# A method for calculating solvation structure on a sample surface from a force curve between a probe and the surface: One-dimensional version

**Ken-ich Amano**

*Department of Chemistry*, *Faculty of Science*, *Kobe University*, *Nada-ku*, *Kobe 657-8501*, *Japan*

Author to whom correspondence should be addressed: Ken-ichi Amano.

Electric mail: k-amano@gold.kobe-u.ac.jp

**MAIN TEXT**

Recent surface force apparatus (SFA) [1] and atomic force microscopy (AFM) [2,3] can measure force curves between a probe and a sample surface in solvent. The force curve is thought as the solvation structure in some articles, because its shape is generally oscilltive and pitch of the oscillation is about the same as diameter of the solvent. However, it is not the solvation structure. It is only the force between the probe and the sample surface. Therefore, this brief paper presents a method for calculating the solvation structure from the force curve. The method is constructed by using integral equation theory, a statistical mechanics of liquid (Ornstein-Zernike equation coupled by hypernetted-chain closure) [4]. This method is considered to be important for elucidation of the solvation structure on a sample surface.

Here, we explain the method for one-dimensional version. The method takes advantage of Ornstein-Zernike (OZ) equation and hypernetted-chain (HNC) closure. Solvent is modeled as an ensemble of small spheres. The small spheres are hard spheres or Lennard-Jones (LJ) spheres. The LJ spheres mean that each sphere interacts with LJ potential (e.g., Octamethylcyclotetrasiloxane (OMCTS) and carbon tetrachloride ($CCl_4$) are one of the typical LJ liquids.). The probe is modeled as a sufficiently large sphere or a small sphere. The former and latter correspond to the probe models of SFA and AFM [2], respectively. The probe is also a hard or LJ sphere. When AFM is considered, the small sphere of the probe model corresponds to a tip of the probe, and effective diameter of which is set according to the sample surface resolution of the AFM tip. In this method, only the sample whose surface is (almost) flat is treated, and hence a huge sphere is presumed as the sample model. However, an infinitely large sphere cannot be applied in computational calculation. Therefore, the sample is modeled as a sufficiently large sphere. The sample is also a hard or LJ sphere. The effective diameters of the models of the solvent, probe, and sample are $d_S$, $d_P$, and $d_M$, respectively.

Before explanation of the method for calculation of the solvation structure, we introduce basic OZ equations and HNC closure. We notify that the former is exact and the latter is approximation. The OZ equation for multi-components is expressed as

$$w_{ij}(r) = h_{ij}(r) - c_{ij}(r) = \sum_m \rho_m \int c_{im}(|\mathbf{r} - \mathbf{r}'|) h_{mj}(r') d\mathbf{r}' \qquad (1)$$

Here, $w = h - c$, $h$ is the total correlation function, $c$ the direct correlation function, and $\rho_m$ the number density of the solvent (solute) of species "m". For example, $h_{ij}$ means the total correlation function between species $i$ and $j$. Since the probe and sample with spherical models are immersed in the solvent in the present case, their correlations can be expressed within Eq. (1). Subscripts $i$ and $j$ indicate the solvent (S), probe (P), and sample (M). "m" also indicates them. Then, $\rho_S$ is number density of the solvent which has a certain value. On the other hands, $\rho_P$ and $\rho_M$ are asymptotically zero. Representation of Eq. (1) can be simplified in the Fourier space using convolution integral as follows,

$$W_{ij}(k) = H_{ij}(k) - C_{ij}(k) = \sum_m \rho_m C_{im}(k) H_{mj}(k), \qquad (2)$$

where capital letters $W$, $H$, and $C$ represent three-dimensional Fourier transforms of $w$, $h$, and $c$, respectively. For example, $W(k)$ is expressed as follows,

$$W(k) = \int w(r) \exp(-i\mathbf{k} \cdot \mathbf{r}) d\mathbf{r} = 4\pi \int_0^\infty r^2 w(r) [\sin(kr)/(kr)] dr. \qquad (3)$$

In Eq. (3), i and π are imaginary unit and circle ratio, respectively. Next, the HNC closure is written as

$$g_{ij}(r) = h_{ij}(r) + 1 = \exp[-u_{ij}(r)/(k_B T) + h_{ij}(r) - c_{ij}(r)], \qquad (4)$$

where $g$ is the reduced density profile (radial distribution function), $u$ the potential, $k_B$

the Boltzmann constant, and $T$ the absolute temperature. The solvation structure on the sample, the goal of this calculation, is expressed as $g_{MS}$. Here, the method for calculating the solvation structure is explained below. A merit of this method is that data of $w_{MP}$ at the overlapped area (between the sample and probe) are not required, which is helpful to SFA and AFM because they cannot measure certain values at the overlapped area.

**[Step 1]** Calculate mean force potential ($\Phi$) from a force curve ($f$) by using following equation,

$$\Phi(r) = \int_r^\infty f(r')dr' \qquad \text{where } r \geq (d_M+d_P)/2. \tag{5}$$

$r$ and $r'$ are distance between the centers of the sample and probe and $(d_P+d_M)/2$ is the contact point. Since the force curve is measured between not models of the sample and probe but actual ones, $f(r')$ is feasibly made from the raw data. (It is made by fitting the contact point between the sample and probe.)

**[Step 2]** Extract solvent mediated potential ($\varphi$) from the mean force potential.

$$\varphi(r) = \Phi(r) - u_{MP}(r) \qquad \text{where } r \geq (d_M+d_P)/2. \tag{6}$$

**[Step 3]** Obtain $w_{MP}$ from $\varphi$.

$$w_{MP}(r) = -\varphi(r)/(k_B T) \qquad \text{where } r \geq (d_M+d_P)/2. \tag{7}$$

**[Step 4]** Calculate $C_{PS}$ by using Eqs. (2) and (4).

**[Step 5]** Introduce a trial function of the total correlation function between the sample and solvent ($h'_{MS}$). Two examples of the trial functions for hard sphere solvent are as follows:

$$h'_{MS}(r) = -1 \qquad \text{for } r < (d_M+d_S)/2 \qquad (8a)$$

$$h'_{MS}(r) = a_1\cos(2\pi a_2 r)\exp(-2\pi a_3 r) \qquad \text{for } r \geq (d_M+d_S)/2, \qquad (8b)$$

$$h'_{MS}(r) = -1 \qquad \text{for } r < (d_M+d_S)/2 \qquad (9a)$$

$$h'_{MS}(r) = a_1\cos(2\pi a_2 r)\exp(-2\pi a_3 r)\sqrt{2/[(1+2\pi r)\pi]}/(1+2\pi r)$$
$$\text{for } r \geq (d_M+d_S)/2. \qquad (9b)$$

Equations (8) and (9) are fabricated empirically, and Eq.(9) is made by referring the Bessel function ($J_{-1/2}$). The trial function for LJ solvent can be introduced in the similar way. $a_1$, $a_2$, and $a_3$ in Eqs. (8) and (9) are variable coefficients. As shown in some articles of SFA and AFM in liquid, the measured force curves are not always similar to the typical oscillative shape. This fact is true of the theoretical results (computational results) [5]. However, solvation structures on diverse surfaces (i.e., solvophobic, neutral, and solvophilic surfaces) always possess the typical oscillative shape in the case of simple liquid. Therefore, Eqs. (8), (9), and the similar trial functions meet the need for the calculation. In this method, the trial function of the direct correlation function ($c'_{MS}$) is not introduced, but that of the total correlation function ($h'_{MS}$) is applied. Although $h'_{MS}$ can be calculated from $c'_{MS}$, $h'_{MS}$ should be introduced as the trial function. This is because, if a trial function of the direct correlation function is used, values of $c'_{MS}$ at the overlapped area (between the sample and solvent) must be estimated, which do not have similar shape in all of the case. Therefore, the trial function of the total correlation function is introduced here.

**[Step 6]** Do the three-dimensional Fourier transform to the trial function of $h'_{MS}$ and output $H'_{MS}$ (see Eq. (3)).

**[Step 7]** Calculate $W'_{MP}$ from $H'_{MS}$ and $C_{PS}$ by using the OZ equation.

$$W'_{MP}(k) = \rho_S H'_{MS}(k) C_{PS}(k). \tag{10}$$

**[Step 8]** Do the three-dimensional inverse Fourier transform to the $W'_{MP}$ and output $w'_{MP}$.

$$w'_{MP}(r) = [1/(2\pi)^3] \int W'_{MP}(k) \exp(i\mathbf{k} \cdot \mathbf{r}) \, d\mathbf{k}$$
$$= [1/(2\pi^2)] \int_0^\infty k^2 W'_{MP}(k)[\sin(kr)/(kr)] dk. \tag{11}$$

**[Step 9]** Integrate differences between the $w'_{MP}$ and $w_{MP}$ by using a formula below,

$$\omega = \int_{(d_M+d_P)/2}^\infty [w'_{MP}(r) - w_{MP}(r)]^2 dr. \tag{12}$$

**[Step 10]** When the $\omega$ is sufficiently small (close to zero), the trial function of $h'_{MS}$ with the variable coefficients is the answer ($h_{MS}$), while in the other situation, return to Step 5 and set the new coefficients into the trial function.

Those Steps 1-10 are the method for one-dimensional version. Using the computational calculation, we have already verified the method and we conclude that the method is applicable.

Once the solvation structure on the sample surface ($g_{MS}$) is calculated, the direct correlation function ($c_{MS}$) can be calculated by using the OZ equation. Thus, by using $h_{MS}$ and $c_{MS}$, the potential between the sample and solvent ($u_{MS}$) can also be calculated, which is written as

$$u_{MS}(\mathbf{r})/(k_B T) = -\ln[h_{MS}(r) + 1] + h_{MS}(r) - c_{MS}(r). \tag{13}$$

Here, Eq. (13) is derived from the HNC closure. The output of $u_{MS}$ can be used for checking of the accuracy of the calculated solvation structure ($g_{MS}$).

We predict that the method for three-dimensional version could also be done by following the same steps, although the calculation time is much longer than the one-dimensional one. A difficult point of the three-dimensional version exists in the trial function, because it should have and deal with many variable coefficients. The number of coefficients should be decreased as much as possible to terminate the calculation within proper time length. A probable countermeasure is introduction of a steepest descent method or a Monte Carlo method into the sampling of the variable coefficients.

**ACKNOWLEDGEMENTS**

We greatly thank Masahiro Kinoshita (Kyoto University) for helpful advises, and appreciate discussions with Hiroshi Onishi (Kobe University) and Kenjiro Kimura (Kobe University). This work was supported by Grant-in-Aid for JSPS (Japan Society for the Promotion of Science) fellows and Foundation of Advanced Technology Institute.

1st Submission: Fri., 31 Aug. 2012 05:07:03 EST (339kb)

2nd Submission: Title is modified and references are added; Sat., 1 Sep. 2012 EST.

3rd Submission: Signs in Eq. (13) are corrected; A sentence is added in the acknowledgements; Mon., 3 Sep. 2012 EST.

4th Submission: A subscript symbol in page 3 is corrected; Fri., 1 Feb. 2013 EST.

5th Submission: A foundation is added to Acknowledgements; Wed., 6 Mar. 2013 EST.